\documentstyle[12pt,epsfig]{article}
\catcode`\@=11
\def\section{\@startsection {section}{1}{0pt}{-3.5ex plus -1ex minus
 -.2ex}{2.3ex plus .2ex}{\raggedright\large\bf}}
\catcode`\@=12
\def\R{{\rm I\!R}}
\def\Q{{\mathchoice
 {\setbox0=\hbox{$\displaystyle\rm Q$}\hbox{\raise 0.15\ht0\hbox to0pt
 {\kern0.4\wd0\vrule height0.8\ht0\hss}\box0}}
 {\setbox0=\hbox{$\textstyle\rm Q$}\hbox{\raise 0.15\ht0\hbox to0pt
 {\kern0.4\wd0\vrule height0.8\ht0\hss}\box0}}
 {\setbox0=\hbox{$\scriptstyle\rm Q$}\hbox{\raise 0.15\ht0\hbox to0pt
 {\kern0.4\wd0\vrule height0.7\ht0\hss}\box0}}
 {\setbox0=\hbox{$\scriptscriptstyle\rm Q$}
 \hbox{\raise 0.15\ht0\hbox to0pt
 {\kern0.4\wd0\vrule height0.7\ht0\hss}\box0}}}}
\def\C{{\mathchoice
 {\setbox0=\hbox{$\displaystyle\rm C$}\hbox{\hbox to0pt
 {\kern0.4\wd0\vrule height0.9\ht0\hss}\box0}}
 {\setbox0=\hbox{$\textstyle\rm C$}\hbox{\hbox to0pt
 {\kern0.4\wd0\vrule height0.9\ht0\hss}\box0}}
 {\setbox0=\hbox{$\scriptstyle\rm C$}\hbox{\hbox to0pt
 {\kern0.4\wd0\vrule height0.9\ht0\hss}\box0}}
 {\setbox0=\hbox{$\scriptscriptstyle\rm C$}\hbox{\hbox to0pt
 {\kern0.4\wd0\vrule height0.9\ht0\hss}\box0}}}}
\font\fivesans=cmr5
\font\sevensans=cmr7
\font\tensans=cmr10
\newfam\sansfam
\textfont\sansfam=\tensans\scriptfont\sansfam=
 \sevensans\scriptscriptfont
\sansfam=\fivesans
\def\sans{\fam\sansfam\tensans}
\def\Z{{\mathchoice
 {\hbox{$\sans\textstyle Z\kern-0.4em Z$}}
 {\hbox{$\sans\textstyle Z\kern-0.4em Z$}}
 {\hbox{$\sans\scriptstyle Z\kern-0.3em Z$}}
 {\hbox{$\sans\scriptscriptstyle Z\kern-0.2em Z$}}}}
\newcommand{\beq}{\begin{equation}}
\newcommand{\eeq}{\end{equation}}
\newcommand{\bea}{\begin{eqnarray}}
\newcommand{\eea}{\end{eqnarray}}
\begin{document}
\begin{titlepage}
\begin{center}
{\Large  Improved actions, the perfect action, and \\
scaling by perturbation theory in Wilsons renormalization \\
group: the two dimensional $O(N)$-invariant non linear \\
$\sigma$-model in the hierarchical approximation} \\[10mm]
\end{center}
\begin{center}
{\Large C. Wieczerkowski $^1$ and Y. Xylander $^2$} \\[10mm]
\end{center}
\begin{center}
$^1$ Institut f\"ur Theoretische Physik I,
Universit\"at M\"unster, \\
Wilhelm-Klemm-Stra\ss e 9, D-48149 M\"unster, \\
wieczer@yukawa.uni-muenster.de \\[2mm]
$^2$ Institut f\"ur Theoretische Physik II,
Universit\"at Hamburg, \\
Luruper Chaussee 149, D-22761 Hamburg \\
xylander@x4u2.desy.de
\end{center}
\vspace{-10cm}
\hfill MS-TP1-94-14
\vspace{11cm}
\begin{abstract}
We propose a method using perturbation theory
in the running coupling constant and the
idea of scaling to determine improved actions
for lattice field theories combining
Wilson's renormalization group with
Symanzik's improvement program .
The method is based on the analysis of a single
renormalization group transformation.
We test it on the hierarchical $O(N)$ invariant
$\sigma$ model in two dimensions.
\end{abstract}
\end{titlepage}

\section{Introduction}
Improved actions for lattice field theories were
proposed by Symanzik \cite{S}
to reduce systematic errors
due to a finite lattice spacing.
The improvement parameters are calculated by
bare perturbation theory.
Improved actions also naturally arise in the
block spin renormalization group approach
of Wilson \cite{WK}.
There they come in form of a renormalized trajectory
of lattice actions completely free of lattice artefacts.
One way to approach the renormalized trajectory
starting from any reasonable ansatz is to iterate
renormalization group transformations.
A different way due to Hasenfratz and Niedermayer
\cite{HN} for asymptotically
free theories consists of a classical treatment
of a single step. It gives the asymptotic form of
the renormalized trajectory called perfect action.
We propose to approximate systematically the
renormalized trajectory in the spirit of Symanzik
by a scaling analysis of a single renormalization group
transformation in terms of perturbation theory in
the running coupling constant.

Consider the renormalization group flow of
an asymptotically free
lattice field $\phi$ on a unit lattice with action
${\cal S}(\phi)$ generated by a renormalization
group transformation ${\rm R}$.
Imagine a one dimensional renormalized trajectory
${\cal S}^{RT}(\phi,f)$
parametrized by the value of a running coupling $f$.
A block spin transformation will do
nothing to an action on this renormalized trajectory
but change the value of the running coupling $f$ into $f^\prime$.
The whole dynamics of the renormalization group
is then encoded in this change of the running coupling.
($f$ can for instance be marginally relevant, which means
that $f^\prime=f+\bigtriangleup\beta (f)$ with
$\bigtriangleup\beta (f)=\beta_2 f^2+O(f^3)$ where $\beta_2>0$.)
An action on the renormalized trajectory therefore
satisfies the discrete Callan-Symanzik type equation
${\rm R}{\cal S}^{RT}(\phi,f^\prime)=S^{RT}(\phi,f)$
and is said to scale.
Recall that we formulate everything in terms of
transformations on unit lattices.
(The reader who prefers to see explicit dependence on
lattice spacings is invited to rewrite the
equations using dimensional analysis.)

In the domain where the running coupling is small,
perturbation theory can be applied to compute
$S^{RT}(\phi,f)$ (at least in a small field
region) as a solution to the Callan-Symanzik type
renormalization group equation. The strategy is as follows.
Consider a lattice action of some general form
${\cal S}(\phi)=\sum_{a} {\rm P}_a(f) {\cal O}_a(\phi)$.
Here ${\rm P}_a(f)$ denote polynomials in the running coupling
containing free improvement parameters.
In the renormalization group terminology the action can
contain irrelevant terms.
We then compute a block spin transformation to a given
order $s$ of perturbation theory in
the running coupling $f$ to obtain
${\rm R}{\cal S}(\phi,f)={\rm R}{\cal S}^{(s)}(\phi,f)+
O(f^{s+1})$. From the result we determine
the effective running coupling $f^\prime$ as a function of $f$,
invert this relation, and express the effective
action in terms of $f^\prime$.
We can then compare the original action as a function of
$f$ with the effective action as a function of $f^\prime$.
The idea is then to demand that the action be invariant up
to corrections of order $s+1$ to determine
the form of the action and fix the values of the
improvement parameters.
Imagine that we have succeded in doing so.
The result is the exact renormalized trajectory
$S^{(s)}(\phi,f)$
up to order $s$ in the running coupling.
In other words we have removed all scaling violations
to a given order in the running coupling.

In practice the improvement procedure is
performed step by step in perturbation theory.
To get started we can take an unimproved action
and see what kind of terms are generated in
a first or second order calculation.
Then the action is extended by the corresponding
terms with couplings depending appropriately
on the running coupling with undetermined coefficients.
Eventually this procedure has to be iterated until no further
terms appear. Fortunately a finite order calculation
can only produce a finite number of terms.
Then the coefficients are fixed as described above
completing the first improvement step.

In a pilot study we have worked out this improvement
program for the hierarchical two dimensional
$O(N)$-invariant $\sigma$-model.
The existence of a renormalized trajectory in this
model and asymptotic freedom (for $N>1$) have been proved
rigorously by Gawedzki and Kupiainen \cite{GK}.
We have no doubts that our action can be proved
to be an excellent approximation to the true
renormalized trajectory in a small field
region using \cite{GK} kind of bounds .
As expected the leading term of the renormalized
trajectory can be obtained
by a perfect action calculation \`a la \cite{HN}.
We take this perfect action as the starting point
of the improvement program.
The hierarchical $\sigma$-model itself is of marginal
interest. However, it is an optimal laboratory
to test renormalization group ideas.
The implementation of our program to the full
$O(N)$-invariant $\sigma$-model is under investigation.

Let us mention that the improvement also works when
dealing with more than one coupling to be renormalized.
The renormalization group is then reduced to a
transformation of a set of running couplings,
the Callan-Symanzik type equation being exactly the one
above. It is conceivable that the improvement program
admits a nonperturbative formulation in terms of
polymer representations.

\section{The hierarchical $O(N)$ model}
In the hierarchical (local approximation)
renormalization group for general $N$ component models
we are left to deal with functions $Z(\phi)$
of a single variable $\phi$ with values in $\R^N$.
For $O(N)$ models one assumes further that
$Z(M\phi)=Z(\phi)$ for all $M\in O(N)$, requiring
$Z(\phi)$ to be a function of $|\phi|$.
We  consider the hierarchical
renormalization group transformation in
two dimensions given by
(the case $L=\sqrt{2}$ of \cite{GK})
\beq
{\rm R}Z(\psi)=
{\cal N}\int{\rm d}\mu_{\gamma}(\zeta)
Z(\psi+\zeta)^2.
\eeq
It preserves $O(N)$-invariance.
$\gamma$ is a real positive number and
\beq
{\rm d}\mu_{\gamma}(\zeta)=
(2\pi\gamma)^{\frac{-N}2} {\rm d}^N \zeta
\exp\left(-\frac{\zeta^2}{2\gamma}\right)
\eeq
the corresponding Gaussian measure on $\R^N$.
${\cal N}$ is an optional normalization constant.
Concerning background material on
hierarchical models see \cite{GK} and
references therein. In the following we will write
\beq
Z(\phi)=\exp\left(-V(\phi)\right).
\eeq
A sufficiently general form of $V(\phi)$ for $\sigma$ models
close to the renormalized trajectory proves to be
\beq
V(\phi)=\sum_{a\geq 2} P_a(f)(|\phi|-\frac1f)^a
\eeq
with $P_2(f)=\frac1{4\gamma}+O(f^2)$ and
$P_a(f)=O(f^a)$ for $a>2$.
The running coupling is given by the inverse radius.
In this setup $\phi$ is not restricted to take values
on the sphere with radius $\frac1f$,
the reason being that our recursion does not preserve
this condition.
The normalization constant ${\cal N}$ is conveniently chosen such
that $V(\phi)=0$ for $|\phi|=\frac1f$.
We adopt this renormalization condition.
When computing single renormalization group
transformations we will speak of the previous action
as the bare action and of the outcome as the
effective or renormalized action.

\section{Perfect action}
To calculate the perfect action \cite{HN} for the
hierarchical $O(N)$ model we write the potential in
the form
\beq
V(\phi)=r^2 {\cal V}(\frac{|\phi|}{r}-1)
\eeq
with $r=\frac1f$. The hierarchical transformation then
takes the form
\beq
\exp\left(-r^{\prime2} {\cal V}(
\frac{1}{r^\prime}|\psi|-1)\right)=
{\cal N}\left(\frac{2\pi\gamma}{r^2}
\right)^{-\frac{N}{2}}
\int {\rm d}^N\xi \exp\left(-
\frac{r^2\xi^2}{2\gamma}
-2r^2{\cal V}(|\frac{1}{r}\psi+\xi|-1)\right).
\eeq
The fluctuation field has been rescaled by $r$.
It turns out in perturbation theory that the effective
radius is
\beq
r^\prime=r-\frac{\gamma}{2}(N-1)\frac{1}{r}+
O(\frac{1}{r^3}).
\eeq
In the limit when $r$ and hence also $r^\prime$ is sent to
infinity the fluctuation integral can be evaluated
by the saddlepoint method.
Rescaling also the block spin $\psi$ by $r$ the equation
for the perfect action becomes
\beq
{\cal V}^{RT}(|\psi |-1)=
\inf_{\xi\in\R^N}\left(
\frac{1}{2\gamma}\xi^2+2
{\cal V}^{RT}(|\psi+\xi |-1)\right).
\eeq
This equation can be solved by the ansatz
$c_2 (|\phi|-r)^2$ with a single parameter $c_2$.
The solution is $c_2=\frac{1}{4\gamma}$.
The perfect action approximation for the
renormalized trajectory in this model is then
given by
\beq
V^{RT}(\phi)=
\frac{r^2}{4\gamma}
(\frac{1}{r}|\phi|-1)^2,
\eeq
The right way to think of this formula is as
a line of fixed points parametrized by $r$ of
the classical renormalization group transformation.
Note that unlike \cite{HN} the action is not
just multiplied by $r^2$.
We have tested this approximation
numerically as will be explained below.

\section{Perturbation theory}
The perturbation expansion for the $O(N)$ model
can be computed to high orders using computer algebra.
Let us explain the method in a second order calculation
for the perfect action. As bare potential we take
\beq
V(\phi)=
\frac{r^2}{4\gamma}
(\frac{1}{r}|\phi|-1)^2.
\eeq
The effective potential will be $O(N)$-invariant.
Without loss of generality we can therefore
take the block spin to be given by
$\psi=(r+\Psi)\hat{e}$ with $\hat{e}$ an $N$ component
unit vector, say $(0,\ldots,0,1)^T$.
The shift of $\Psi$ serves to place us into the
minimum of the bare potential.
We then decompose orthogonally the fluctuation field into
$\zeta=\sigma\hat{e}+\pi$ with respect to the
direction of $\psi$. The one component variable
$\sigma$ is the radial fluctuation field
while the $N-1$ component variable
$\pi$ is the tangential fluctuation field.
The bare potential is expanded in powers of
the coupling $f=\frac1r$. Up to second order
it is given by
\bea
V(\psi+\zeta)&=&
\frac{1}{4\gamma}\Psi^2+
\frac{1}{2\gamma}\Psi\sigma+
\frac{1}{4\gamma}\sigma^2+
\frac{f}{4\gamma}\Psi\pi^2+
\frac{f}{4\gamma}\sigma\pi^2- \nonumber \\
& &\frac{f^2}{4\gamma}\Psi^2\pi^2-
\frac{f^2}{2\gamma}\Psi\sigma\pi^2-
\frac{f^2}{2\gamma}\sigma^2\pi^2+
\frac{f^2}{16\gamma}(\pi^2)^2+
O(f^3).
\eea
One immediately observes a linear and a quadratic term in
$\sigma$ which cannot be treated as perturbations.
The mass term changes the $\sigma$ covariance
to $\frac12 \gamma$. The source term is taken into account
by a shift of the radial variable
$\sigma=\xi-\frac12 \Psi$.
The renormalization group transformation then becomes
\beq
\exp \left(-{\rm V}^\prime(\Psi)\right)=
\exp \left(-\frac1{4\gamma}\Psi^2\right)
{\cal N}^\prime
\int {\rm d}\mu_{\frac12\gamma}(\xi)
\int {\rm d}\mu_{\gamma}(\pi)
\exp \left(-2{\rm V}(\Psi,\xi,\pi)\right)
\eeq
in terms of $\xi$ and $\pi$.
The bare potential takes the form
\bea
{\rm V}(\Psi,\xi,\pi)&=&
\frac{f}{8\gamma}\Psi\pi^2+
\frac{f}{4\gamma}\xi\pi^2- \nonumber \\
& &\frac{f^2}{16\gamma}\Psi^2\pi^2-
\frac{f^2}{4\gamma}\Psi\xi\pi^2-
\frac{f^2}{4\gamma}\xi^2\pi^2+
\frac{f^2}{16\gamma}(\pi^2)^2+
O(f^3).
\eea
At this point perturbation theory is applicable.
Although the potential is non-polynomial to begin with
only finitely many terms show up at finite order
with a leading trilinear vertex.
Note that the perfect action is recovered when
fluctuations are completely neglected.
The perturbation expansion is straight forward
using the Gaussian correlations
\beq
\int {\rm d}\mu_{\frac12\gamma}(\xi)
\xi^{2n}=\left(\frac{\gamma}{2}\right)^n
\prod_{m=0}^{n-1}(2m+1)
\eeq
and
\beq
\int {\rm d}\mu_{\gamma}(\pi)
(\pi^2)^n=\gamma^n
\prod_{m=0}^{n-1}(2m+N-1).
\eeq
Computing the fluctuation integral to second order
perturbation theory we obtain an effective potential
of the form
\beq
{\rm V}^\prime(\Psi)=
\left(\frac{1}{4\gamma}-\frac{3}{16}(N-1)f^2\right)\Psi^2
+\frac{1}{4}(N-1)f\Psi
+O(f^3).
\eeq
We then determine the value $\delta r$ of
$\Psi$ at which the effective potential attains its
minimum and substitute $\Psi=\Phi+\delta r$.
The change of $r$ is due to the linear term in $\Psi$.
The meaning of this variable is
$\Phi=|\psi |-r^\prime$ with $r^\prime=r+\delta r$
the renormalized radius.
To second order perturbation
theory in $f$ the change of the radius is
$\delta r=-\frac{\gamma}{2}(N-1)f+O(f^3)$.
{}From this we find a renormalized coupling
$f^\prime=\frac1{r^\prime}$ of the form
\beq
f^\prime=f+\frac{\gamma}{2}(N-1)f^3+O(f^5).
\eeq
(The vanishing of the $f^4$ term of the
$\bigtriangleup\beta$ function follows from a
fourth order calculation.)
In particular we have confirmed that the model is
perturbatively asymptotically free for $N>1$.
That is, when perturbation theory applies we find
a flow where the coupling slowly grows and
the radius slowly shrinks.
The effective potential becomes
\beq
{\bf V}^\prime(\Phi)=
\left(\frac{1}{4\gamma}-\frac{3}{16}
(N-1)f^{\prime 2}\right) \Phi^2
+O(f^{\prime 3})
\eeq
in terms of $\Phi=|\psi |-r^\prime$ and $f^\prime$.
We also see that this action remains invariant in
the sense of scaling to first order.
Scaling violation shows up in a second order flow
of the overall prefactor. In a zeroth improvement step
they can be compensated for by changing the
bare action into
\beq
V(\phi)=
\left( \frac{1}{4\gamma}+c_2^{(2)} f^2\right)
\left(|\phi|-\frac{1}{f}\right)^2.
\eeq
The correct value of the improvement parameter is
$c_2^{(2)}=\frac{3}{8}(N-1)$.
The resulting action can then be seen to scale
even to second order.

To appreciate this result consider
for instance the polynomial bare potential
\beq
V(\phi)=
\frac{r^2}{\gamma}\left(
\frac{1}{r^2}\phi^2-1\right)^2.
\eeq
Perturbation theory reveals that its effective action
contains among other terms a cubic one at first order
in the effective coupling. Therefore this action does
not scale even to first order.

\section{Improved action}
Let us now also remove the scaling violations of
second order. As a bare potential we take the second order
improved one which we write in the form
\beq
 V^{(2)}(\phi)=P_2^{(2)}(f)
\left(|\phi|-\frac{1}{f}\right)^2
\eeq
with
$P_2^{(2)}(f)=\left[\frac{1}{4\gamma}+
\frac{3}{8}(N-1) f^2\right]$.
The effective potential computed to third order perturbation theory
is given by
\beq
{\bf V}^{\prime(2)}(\Phi)=
\left(\frac{1}{4\gamma}-\frac{3}{8}
(N-1)f^{\prime 2}\right) \Phi^2+
\frac{7}{48}(N-1)f^{\prime 3} \Phi^3
+O(f^{\prime 4})
\eeq
in terms of $f^{\prime}$. What is new is a cubic term
in $\Phi=|\phi|-r^\prime$. As it will be generated anyway
to third order it is natural to include it to this order
in the bare action. In other words let us make the ansatz
\beq
 V^{(3)}(\phi)=P_2^{(3)}(f)\left(|\phi|-\frac{1}{f}\right)^2
+P_3^{(3)}(f)\left(|\phi|-\frac{1}{f}\right)^3
\eeq
with $P_2^{(3)}(f)=P_2^{(2)}(f)+c^{(3)}_2 f^3$ and
$P_3^{(3)}(f)=c^{(3)}_3 (N-1)f^{\prime 3}$.
Here we have anticipated a possible cubic correction
to $P_2(f)$ which in fact turns out not to exist.
The ansatz contains two new improvement parameters.
Computing again the effective potential to third order,
the correct values of these are seen to be
$c^{(3)}_2=0$ and $c^{(3)}_3=\frac{7}{36}$.
This action reproduces itself
up to fourth order corrections.
The improvement scheme can now be iterated.
Suppose that we have found the potential
\beq
V^{(s)}(\phi)=\sum_{a=2}^{s}
P_a^{(s)}(f)\left(|\phi|-\frac1f\right)^a
\eeq
which scales up to order $s$.
It's image under a renormalization group transformation
evaluated in perturbation theory to order $s+1$ is of the
form
\beq
V^{(s+1)}(\phi)=\sum_{a=2}^{s+1}
P_a^{(s+1)}(f)\left(|\phi|-\frac1f\right)^a
\eeq
containing polynomials
$P_a^{(s+1)}(f) =\left[P_a^{(s)}(f)+c_a^{(s+1)}f^{s+1}\right]$
for $a\leq s$ and
$P_{s+1}^{(s)}(f)=c_{s+1}^{(s+1)}f^{\prime s+1}$.
This general form again is reproduced to this order.
The corresponding effective potential to order $s+1$
is given by
\beq
{\bf V}^{\prime(s+1)}(\Phi)=
\sum_{a=2}^{s+1} P_a^{\prime(s+1)}(f') \Phi^a
+{\cal O}(f^{\prime s+2})
\eeq
with
$P_a^{\prime(s+1)}(f)=\left[P_a^{(s)}(f')
+c_a^{\prime(s+1)} f^{\prime s+1}\right]$
for $a\leq s$ and
$P_{s+1}^{\prime(s+1)}(f) =c_{s+1}^{\prime(s+1)} f^{\prime s+1}$.
The effective coefficients $c^{\prime (s+1)}_a$ depend linearly
on their bare counterparts $c^{(s+1)}_a$.
(To order $s+1$ they have no other choice.)
In order that there be no scaling violation to
order $s+1$ the polynomials $P_a^{\prime(s+1)}(x)$ and
$P_a^{(s+1)}(x)$ have to be equal. From this we obtain a
system of linear equations for the coefficients $c_a^{(s+1)}$.
This system turns out to have a unique solution.
To order $f^6$ it is given by
\bea
P_2^{(6)}(f)&=&
{\frac {1}{4\,\gamma}}-
{\frac {3}{8}\,(N-1)}f^{2}+
\left ({\frac {29\,\gamma\,\left (N-1\right )}{56}}+
{\frac {17\,\gamma\,\left (N-1\right)^{2}}{12}}\right)f^{4}-
\nonumber \\ &&
\left({\frac {571\,\gamma^{2}\left (N-1\right )^{3}}{84}}+
{\frac {327167\,\gamma^{2}\left (N-1\right )^{2}}{70560}}+
{\frac {10305\,\gamma^{2}
\left (N-1\right )}{3472}}\right )f^{6},
\nonumber \\
P_3^{(6)}(f)&=&\frac {7}{36}\,(N-1)f^{3}-
\left({\frac {239\,\gamma\,\left (N-1\right )}{360}}+
{\frac {361\,\gamma\,
\left (N-1\right )^{2}}{336}}\right )f^{5},
\nonumber \\
P_4^{(6)}(f)&=& -{\frac {15}{112}\,(N-1)}f^{4}+
\left({\frac {1479\,\gamma\,\left (N-1\right )}{1736}}+
{\frac {37753\,\gamma
\,\left (N-1\right )^{2}}{35280}}\right )f^{6},
\nonumber \\
P_5^{(6)}(f)&=& {\frac {31}{300}\,(N-1)}f^{5},
\nonumber \\
P_6^{(6)}(f)&=& -{\frac {21}{248}\,(N-1)}f^{6}.
\eea
We observe that the power series for
$P_a(f)$ contains only even (odd) powers of $f$ when
$a$ is even (odd).
Furthermore we observe that the signs of the
coefficients alternate.
The complete series is not expected to converge due
to instanton singularities. It would however be
very interesting to apply the machinery of
resummation methods to a high order approximation
to the renormalized trajectory.
What remains to be said to order $f^6$ is the
recursion formula
\beq
f'=f+{\frac {\gamma\,\left (N-1\right )}{2}}f^{3}-
{\frac {\gamma^{2}\left (N-1\right )}{12}}f^{5}+
O(f^7).
\eeq
for the running coupling. For the sake of completeness
we also include the recursion
\bea
r^\prime&=&r-\frac{\gamma}{2}(N-1)\frac{1}{r}+
\frac{\gamma^2}{12}
\left(3(N-1)^2+N-1\right)\frac{1}{r^3}-
\nonumber \\ &&
\frac{\gamma^3}{48}
\left(18(N-1)^3-11(N-1)^2+32(N-1)\right)\frac{1}{r^5}+
O(f^7)
\eea
for the effective radius.

\section{Numerical results}
\begin{figure}[h]
\centerline{
  \hbox{
  \epsfig{file=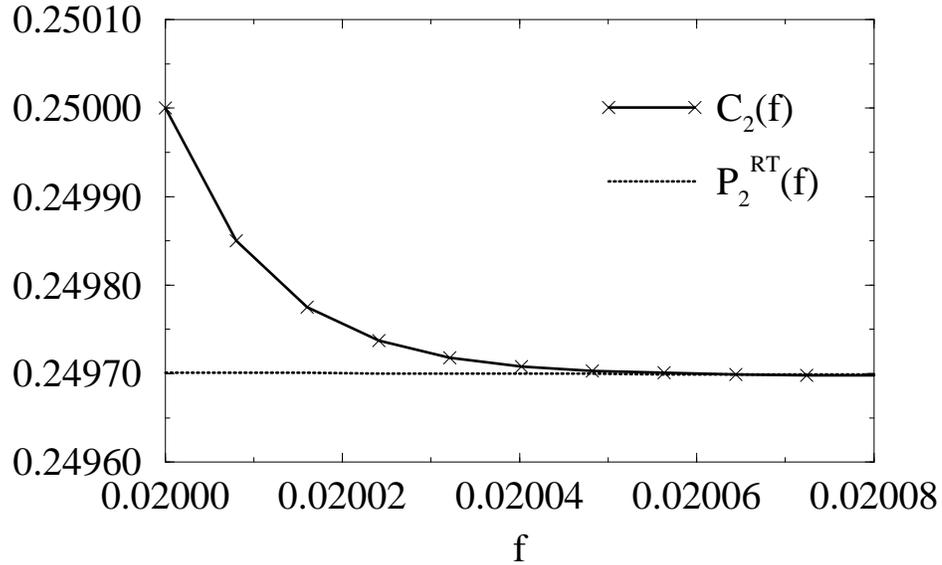,width=12.5cm}
}
}
\centerline{
\parbox{12.5cm}{
\caption[C2FLOW]{Flow of the coupling $C_2(f)$ towards it's
  renormalized value $P_2^{RT}(f)$ under consecutive applications of
the RG transformation. The calculation was performed with a bare radius
of 50. The other couplings  $C_a(f)$ with $2<a\le6$ behave similar.
\label{C2FLOW}
}
}
}
\end{figure}

\begin{figure}[h]
\centerline{
  \hbox{
  \epsfig{file=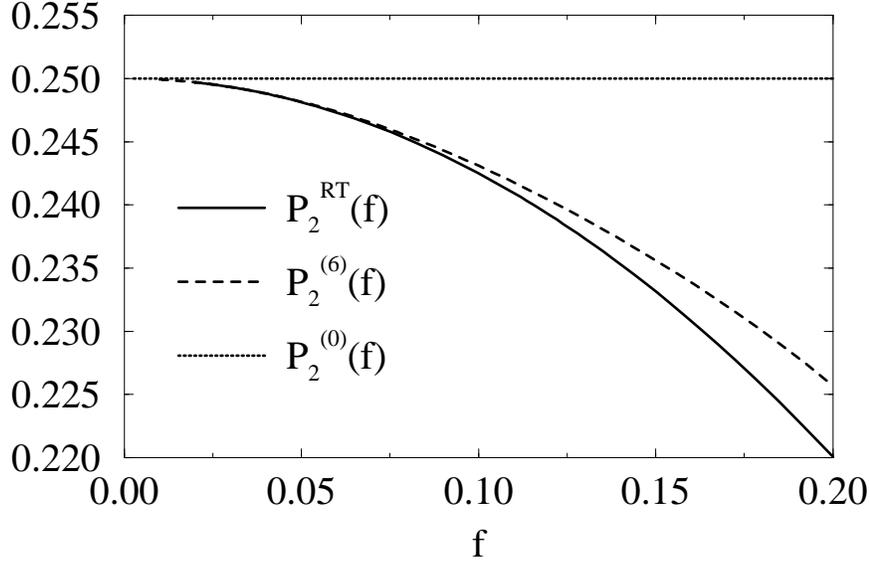,width=12.5cm}
}
}
\centerline{
\parbox{12.5cm}{
\caption[P2RT]{
Comparison of the perfect renormalized coupling $P_2^{RT}(f)$ with
it's improved and unimproved approximations $P_2^{(6)}(f)$,$P_2^{(0)}(f)$
\label{P2RT}
}
}
}
\end{figure}

\begin{figure}[h]
\centerline{
  \hbox{
  \epsfig{file=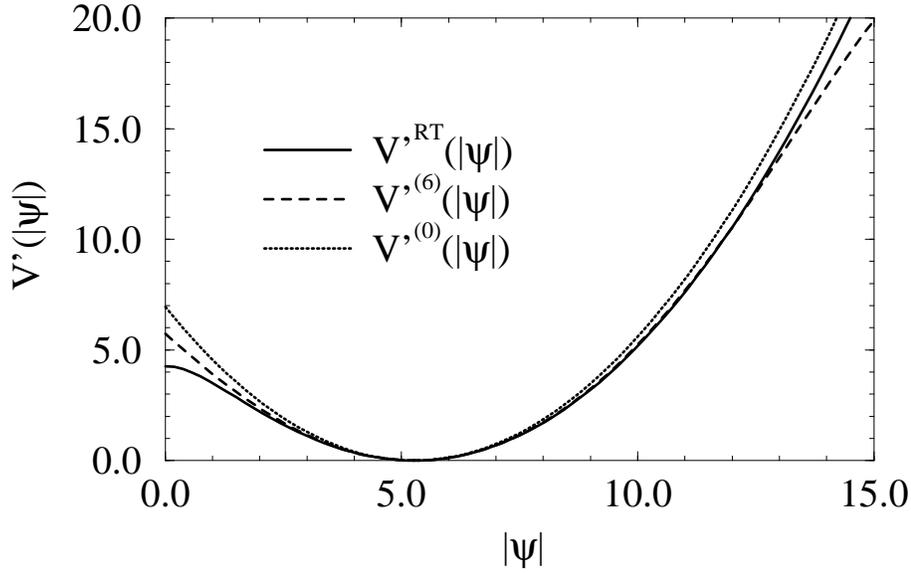,width=12.5cm}
}
}
\centerline{
\parbox{12.5cm}{
\caption[RR]{
Large field behavior of the effective potential.
\label{Large}
}
}
}
\end{figure}

Fortunately we are not left alone with perturbation
theory. Hierarchical renormalization group transformations
can also be computed numerically.
Therefore we are able to test our perturbative
results and find the limits of their validity.
Going back to the point of departure we write the
transformation in the form
\beq
\exp\left(-{\sf V}'(|\psi|)\right)=
{\cal N}\int {\rm d}^N\xi \exp\left(-
\frac{1}{2\gamma}\xi^2-2{\sf V}(|\psi+\xi|)\right).
\eeq
with ${\sf V}(|\phi|)=V(\phi)$.
The fluctuation integral is most conveniently done
in $N$ dimensional polar coordinates given by
$|\xi|=R$ and $\psi \cdot \xi = |\psi|\, R\, \cos\theta$.
The independent polar angles are immediately integrated
over, changing just the (irrelevant) normalization.
The result is
\beq
\exp\left(-{\sf V}'(|\psi|)\right)=
{\cal N'}\int \limits_{0}^{\infty}{\rm d} R
\int \limits_{0}^{\pi}{\rm d} \theta
\;w(R,\theta) B\left(\sqrt{|\psi|^2+2 |\psi| R \cos \theta+
R^2} \right)
\label{numint}
\eeq
with a weight function
\begin{equation}
  w(R,\theta) := R^{N-1}\, \sin^{N-2}\theta
\exp\left(-\frac{1}{2\gamma}R^2\right)
\end{equation}
and a Boltzmann factor (squared by the block volume two)
\begin{equation}
B(\varphi) :=
\exp\left(
-2{\sf V}(\varphi)\right).
\end{equation}
$\varphi$ denotes a positive real valued
single component field (the modulus of $\psi$).
In a numerical evaluation the range of
the $R$-integration can be restricted
to $[0,R_{max}]$ with $R_{max}$ sufficiently large.
The contribution of $(R_{max},\infty )$ is both supressed by
the exponential decay of the weight function $w(R,\theta)$
and the decay of the Boltzmann factor.
Furthermore it is sufficient to follow the flow of
$B(\varphi)$ in an interval
$\varphi \in [\varphi_{min},\varphi_{max}]$
with numerically negligible corrections.
Let us emphasize that the interval should include a
large field region not covered by perturbation theory
in order to test our improved potential.
We parametrize ${\sf V}(\varphi)$ by a set
of $N_{\varphi}$ values ${\sf V}_i={\sf V}(\varphi_i)$
at equidistant points $\varphi_i \in [\varphi_{min},\varphi_{max}]$.
Intermediate values are computed by cubic-spline interpolation.
The integral (\ref{numint}) is performed by standard numerical
methods yielding a set of effective values ${\sf V}^{\prime}_i$
from which we compute parameters $r'$ and $C'_a$ by a
least square fit using the ansatz
\beq
{\sf V}'(\varphi)=\sum_{a=2}^{s} C'_a\left(\varphi-r'\right)^a.
\eeq
The renormalized trajectory can be computed
by iterative renormalization group steps.
(All numerical results refer to parameter values
$\gamma=1,N=3,\varphi_{min}=r-30,\varphi_{max}=r+30,N_\varphi=160$.)
We start with an unimproved bare potential
$V^{(0)}(|\phi|)=1/(4\gamma)(|\phi|-r)^2$
and iterate renormalization group transformations.
After each step the resulting effective potential comes closer
to the renormalized trajectory giving successivly
better approximations (see figure \ref{C2FLOW}).
We took the values of the couplings $C'_a(f')=:P_a^{RT}(f')$
after 10 iterations as estimates of the
true renormalized couplings at the radius $r'=1/f'$.
By varying the bare radius $r$ we computed
the renormalized trajectory for a set of renormalized radii
$r'\in [0,100]$.
The results were then compared
with the coefficients $P_a^{(6)}(f')$ of the improved
perfect action approximation $V^{(6)}$.
As expected one finds excellent agreement
(fig. \ref{P2RT})
for large $r=\frac{1}{f}$.
The relative deviation of the perturbative coefficients
from their nonperturbative values is less than
$10^{-3}$ for $r>30$ and remains smaller than two percent
for $r>5$.
The improvement compared to the unimproved perfect action
is especially visible in the large field region (fig. \ref{Large}).

\begin{figure}[h]
\centerline{
  \hbox{
  \epsfig{file=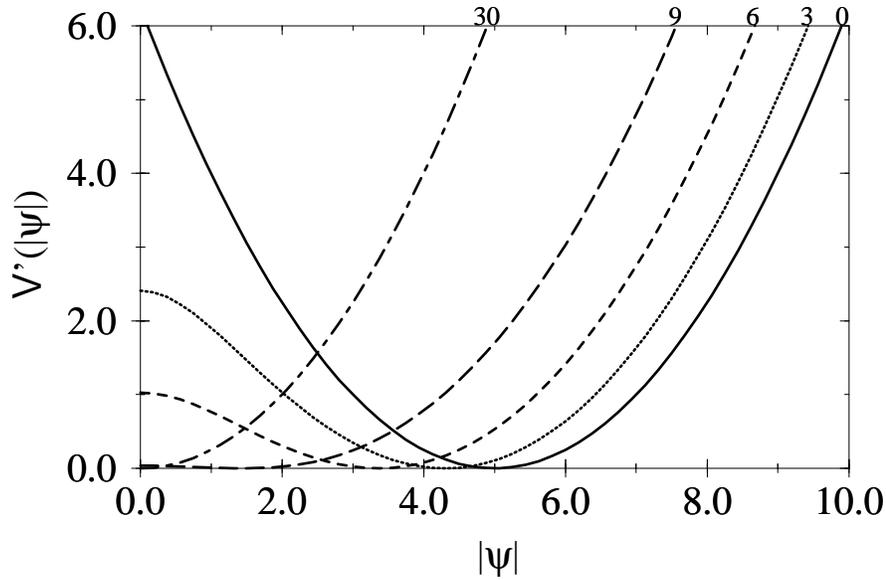,width=12.5cm}
}
}
\centerline{
\parbox{12.5cm}{
\caption[Vflow]{Flow of the effective potential. The plot shows $V'(|\psi|)$
after n=0,3,6,9 and 30 RG steps starting with the unimproved
perfect action approximation $V^{(0)}(|\psi|)=1/(4\gamma)(|\psi|-r)^2$ at $r=5$
and ending in the HT fixpoint
$V_{HT}(|\psi|)=1/(4\gamma)|\psi|^2$.
}
}
}
  \label{Vflow}
\end{figure}

\begin{figure}[h]
\centerline{
  \hbox{
  \epsfig{file=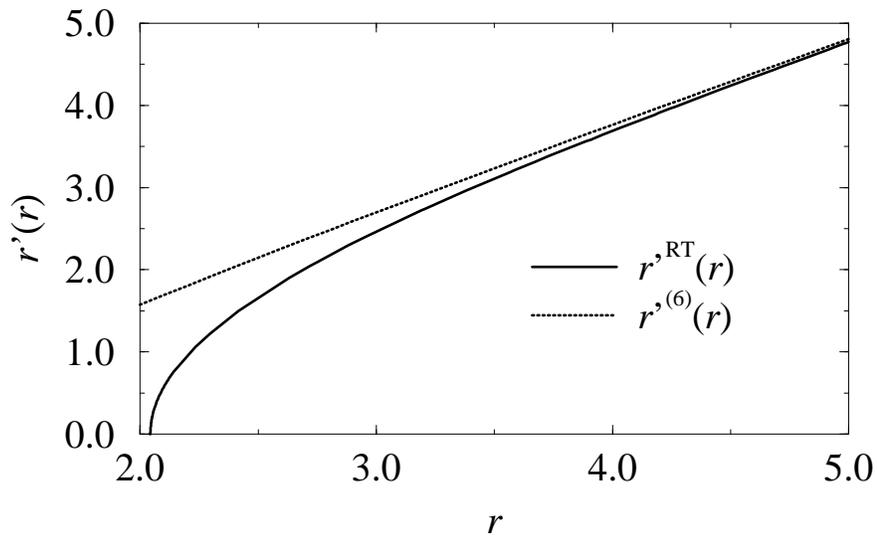,width=12.5cm}
}
}
\centerline{
\parbox{12.5cm}{
\caption[RR]{
Flow of the radius in the IR. At $r_c\approx2.04$ the effective radius
vanishes with $r'\approx|r-r_c|^{0.5}$.
\label{RR}
}
}
}
\end{figure}

Around $r=5$ the deviation increases dramatically.
This breakdown of perturbation theory
in the crossover region from the
ultraviolet to the infrared regime is illustrated in
fig. \ref{Vflow}.
The renormalized trajectory connects the UV-fixpoint
with the IR-fixpoint.
The IR-fixpoint potential is a quadratic single well
whereas the UV-fixpoint potential has a double well shape.
Along the renormalized trajectory
the radius $r$ and the depth  of this well shrink  simultanously.
At some (pseudo) critical point $r_c\approx 2.04$ (see fig. \ref{RR})
the radius
and the depth vanish and remain zero all the way to
the IR fixed point.
Obviously the coupling $f=\frac{1}{r}$ diverges at this
point and looses its meaning as expansion parameter.
Though highly interesting in its own we do not intend
to investigate the infrared behavior in terms of
the flow in this paper. Let us only mention that it is
conceivable that one could performe a change of
coordinates to another expansion parameter in
the crossover region. A natural candidate is the
mass parameter in the potential since the double well
becomes very flat before it turns over.

\section*{Acknowledgment}
We would like to thank M. Grie\ss l for helpful comments.

\vspace{2cm}


\end{document}